\documentclass[preprint,authoryear,12pt]{elsarticle}

\usepackage{amssymb}

\journal{13th Marcel Grossmann Meeting}
\begin{document}

\begin{frontmatter}



\title{The quantum clock: a critical discussion on space-time}


\author[1]{Luciano Burderi}
\author[2]{Tiziana Di Salvo}

\address[1]{Dipartimento di Fisica, 
Universit\`a degli Studi di Cagliari, 
SP Monserrato-Sestu, KM 0.7, 09042 Monserrato, Italy}
\address[2]{Dipartimento di Fisica, 
Universit\`a degli Studi di Palermo, 
via Archirafi 36, 90123 Palermo, Italy}

\begin{abstract}
We critically discuss the measure of very short time intervals.
By means of a {\it gedankenexperiment}, we describe an ideal clock 
based on the occurrence of completely random events. 
We show that the minimum time interval $\Delta t$ that this clock 
can measure scales as the inverse of its size $\Delta r$. This implies an uncertainty 
relation between space and time
$$
\Delta r \Delta t \geq G \hbar / c^{4},
$$ 
where $G$, $\hbar$, and $c$ are the gravitational constant, the reduced Planck 
constant, and the speed of light, respectively. 
We outline and briefly discuss the implications of this uncertainty 
principle.
\end{abstract}


\end{frontmatter}


\section{Introduction}
\label{intro}

The definition of a quantity or a concept in physics has to be 
operational in order to clarify the terms in which that quantity 
should be used and to avoid unjustified assumption of properties 
that belong more to our mental representation of that quantity 
than to its effective nature (e.g.\ Bridgman P. 1927, {\it The Logic of 
Modern Physics}, MacMillan, New York).

This point of view has been particularly fruitful {\it e.g} when 
applied to the critical discussion of the concept of simultaneity, 
leading to the foundation of Special Relativity (SR) 
(Einstein A. 1905, {\it Zur Elektrodynamik bewegter K\"orper}, 
Annalen der Physik, 322, 10). 
Indeed, it is worth noting that an operational definition of 
time is crucial in SR. 
In particular the setting-up of a device that defines time 
in an operational way, whose behavior is constrained by the 
postulate of the invariance of the speed of light, 
implies directly the heterodox phenomenon of time dilation.
Such a device is the so called Light-Clock: two plane parallel
mirrors, facing each other at a distance $dx$, over which a light 
pulse bounces back and forth beating time in equal intervals of
duration $\Delta t = dx/c$, where $c$ is the speed of light.  


In what follows we adopt the rigorously operational 
definition of time as:
\begin{equation} 
\label{eq:timedef}
{\rm time \equiv a\; physical\; quantity\; that\; is\; measured\; 
by\; an\; appropriate\; clock} 
\end{equation}
This apparently trivial (or somewhat circular) definition is essential 
to point out some subtle features of this elusive quantity.

Since it is impossible to measure the position of a particle with 
an accuracy shorter than its reduced Compton wavelength, $\lambda_{\rm C}$
(e.g. Garay L. J., 1995, Int. J. Mod. Phys. A10, 145), 
the minimum distance between the electrons of each of the two mirrors 
over which a light pulse bounces back and forth in a Light-Clock is 
$dx \ge \lambda_{\rm C} = \hbar / (mc)$, where $\hbar$ is the reduced Planck 
constant and $m$ is the particle mass. Thus the shortest time interval such a clock
can measure is $\Delta t = dx/c \ge \hbar / (m c^2)$ which is 
$\simeq 1.3 \times 10^{-21}$ s for electrons. 
Indeed the shortest time interval ever measured is about 20 attoseconds 
$= 2 \times 10^{-17}$ s (Schultze M., {\it et al.} 2010, Delay in 
Photoemission, Science, 328, 1658).

On timescales shorter than $\hbar / (mc^2)$ the Light-Clock fails as a 
device to gauge time and, according to (\ref{eq:timedef}), 
a new device has to be imagined before a physical definition of
time can be extended well below such short intervals.
In the next section we describe an ideal quantum device that is, in 
principle, capable of measuring arbitrarily short time intervals with 
any given accuracy. Curiously, this device is based on a process that, 
in some sense, is just the opposite of a strictly periodic phenomenon, 
namely the (in some respect more fundamental) occurrence of totally 
random events, such as the decay of an ensemble of non-interacting 
particles in an excited state. 
In this case the time elapsed may be obtained by the amount 
of particles that have decayed. Such a device has been discussed in 
Salecker, H., \& Wigner, E. P. (1958, Phys. Review, 109, 571) as an 
example of a simple microscopic clock.
This will result in the construction of an ``ideal'' clock,
the ``Quantum Clock'', that is capable of measuring 
arbitrarily short time intervals with the required accuracy. 
Limits, imposed on our device by Quantum Mechanics (QM) and GR, 
result in a new uncertainty relation that we briefly discuss.

\section{The Quantum Clock}
\label{quantumclock1}

Let us consider a statistical process whose probability of occurrence
\begin{equation} 
\label{eq:random}
dP = \lambda \; dt
\end{equation} 
is independent of time ({\it i.e.} $\lambda$ constant with time).
A good example of this sort of situation is given by radioactive decay.
Given an amount of radioactive matter of mass $M = N m_{\mathrm p}$, 
where $N$ is the number of particles and $m_{\mathrm p}$ is the mass 
of a single particle, it can be 
easily proved that, if equation (\ref{eq:random}) holds, 
the mean variation of the number of particles in the unit time interval is given by
\begin{equation}
\label{eq:decay}
\frac{dN}{dt} = - \lambda N
\end{equation}
(see {\it e.g.} Bevington P. 1969, {\it Data Reduction and Error Analysis 
for the Physical Sciences}, McGraw-Hill, New York).
The mean number of decays in a time $\Delta t << \lambda^ {-1} $ is 
$\Delta N_{\Delta t} = \lambda N \Delta t$. 
The measured number of decays fluctuates around the expected value 
with Poissonian statistics 
{\it i.e} with $\sigma = \sqrt{\lambda N \Delta t}$. 
Therefore it is possible to measure a time interval $\Delta t$ counting 
$\lambda N \Delta t$ 
events. The relative error on our measure is 
$\sigma_t / \Delta t = \epsilon = (\lambda N \Delta t)^{-1/2}$.
As the time intervals become shorter, $\Delta t \rightarrow 0$, 
in order to have a small relative error, say 
$\sigma_t / \Delta t < \epsilon_0$, 
one should keep the product $N \Delta t > (\lambda \epsilon_0^2)^{-1}$. 
Providing that enough particles are available, 
$N$ can be conveniently increased up to the required precision. 

A physical device based on the process discussed above can be built 
in several ways. 
The simplest (albeit perhaps not practical) device consists
of a given amount of mass $M$ of radioactive particles 
(corresponding to a given 
number of particles $M = N m_{\mathrm p}$) completely surrounded by 
proportional counters ({\it e.g.} Geiger--Muller counters) of quantum 
efficiency = 1. We consider this Quantum Clock (QmCl) for 
the operational definition of time. If we count a number of decays 
${\cal N}_{\Delta t} \sim \Delta N_{\Delta t} = \lambda N \Delta t$ 
in the QmCl the time elapsed is
\begin{equation}
\label{eq:1clock}
\Delta t  = \frac{{\cal N}_{\Delta t}m_{\mathrm p} }{\lambda M}
\end{equation}
where we have expressed the number of particles in terms of their mass.
The associated relative accuracy is $\epsilon = \delta t / 
\Delta t \sim {\cal N}_{\Delta t}^{-1/2}$. Because at least one event 
must be recorded by the device we have $ \epsilon \le 1$. 
Therefore in terms of this uncertainty, and
expressing $m_{\mathrm p}$ in terms of its rest energy $E_{\mathrm p} =
m_{\mathrm p} c^{2}$, the time elapsed is
\begin{equation}
\label{eq:2clock}
\Delta t  = \frac{1}{\epsilon^{2} M c^{2}} \times 
\frac{E_{\mathrm p}}{\lambda}.
\end{equation} 

\section{The Quantum Clock and the Heisenberg Uncertainty Principle}
\label{heisenberg}

As a quantum device, the QmCl is subject to the uncertainty relations.
In particular we will use the uncertainty relation between energy and 
time, namely 
\begin{equation}
\label{eq:heis}
\delta E \times \delta t \ge \hbar / 2
\end{equation}
where $\delta E$ and $\delta t$ are the uncertainties in the 
energy and time of a phenomenon.

In (\ref{eq:2clock}) the factor $E_{\mathrm p} / \lambda$ depends 
on the specific
nature of the radioactive substance used for the construction of 
the clock.
To make the QmCl independent of the particular substance adopted, 
we consider the limitations imposed by (\ref{eq:heis}).
Let us suppose that we want to build the clock as light as possible (the
reason for this will be explained in \S \ref{relativity}). To this end we
use a radioactive substance that is completely destroyed in the decay 
process
({\it e.g.} $\pi_{0} \rightarrow 2 \gamma$) otherwise, once the 
decay occurred, the relic particle, that is no more involved in 
further decays,
merely weighs down the clock mass. In this case the energy involved
in the decay process is the whole energy of the particle, {\it i.e.} 
$E_{\mathrm p}$. Conservation of mass-energy imposes an upper limit 
to the
uncertainty in the decay energy $\delta E_{\mathrm p}$ namely
$E_{\mathrm p} \ge \delta E_{\mathrm p \; max}$
where $\delta E_{\mathrm p \; max}$ is the maximum uncertainty obtainable
in the measure of the decay energy.
The decay rate $\lambda$ is the inverse of the average decay time
$1/\lambda = \tau  \ge \delta t_{\mathrm min}$ 
where $\delta t_{\mathrm min}$ is the minimum uncertainty obtainable
in the measure of the time elapsed before the decay.
Since the maximum uncertainty in the energy and the minimum 
uncertainty in the time elapsed are related by the uncertainty 
relation (\ref{eq:heis}), we have:
\begin{equation}
\label{eq:2heis}
\frac{E_{\mathrm p}}{\lambda} = E_{\mathrm p} \times \tau \ge
\delta E_{\mathrm p \; max} \times \delta t_{\mathrm min} \ge \hbar / 2
\end{equation}
Inserting (\ref{eq:2heis}) in (\ref{eq:2clock}) we have
\begin{equation}
\label{eq:3clock}
\Delta t  \ge \frac{\hbar}{2 \epsilon^{2} c^{2}} \times \frac{1}{M}
\end{equation}
which expresses the same lower limit for the mass of a clock, 
capable to measure time intervals down to an accuracy $\Delta t$, 
given by Salecker \& Wigner (1958, see eq. 6 in their paper).
More recently, Ng and van Dam (Ng Y. J., van Dam H., 2003, CQGra, 20, 393, and
reference therein) discussed a similar relation which limits the precision
of an ideal clock (see eq. 8 in their paper).
In the above relation the ``fuzziness'' of the QM manifest itself in the
inequality. However the mass in the denominator of the second member allows,
in principle, to build such a massive clock that an arbitrarily short time
interval can be adequately measured with the required accuracy $\epsilon$.

\section{The Quantum Clock and General Relativity}
\label{relativity}

In GR time is a {\it local} quantity in the sense that
time can flow at different rates in different points of space.
If, for instance, a non-uniform gravitational field is present, the time
flows slower where the gravitational filed is more intense.
Because of its spatial extension, a clock defined by (\ref{eq:3clock}) 
is capable of measuring a sort of ``average'' time interval over the region 
defined by its size. Since we measure time counting events which may
occur randomly in any point of the clock, the size of the region over 
which we are measuring the time is identified with the entire size
of the clock. In other words a spatial uncertainty, corresponding to 
the finite size of the clock, is associated with the measure 
of time. To minimize this uncertainty (providing that the single particles
are so weakly interacting with each other that their behavior is unaffected 
by the proximity of neighbors) it is possible to compress the QmCl 
in order to make its spatial extension as small as possible. 

However the presence of the clock mass affects the structure of space-time. 
In particular GR states that it is {\it impossible} to avoid complete 
gravitational collapse if a given amount of matter $M$ is compressed 
into a volume smaller than its event horizon. 
In the following, we will assume, for simplicity, that this volume is 
spherically symmetric and coincident with a sphere whose radius is the
Schwarzschild radius associated to that amount of matter.
Thus we assume:
\begin{equation}
\label{eq:Schwarzschild}
R_{\rm Sch} = \frac{2 G M}{c^2} 
\end{equation}
where $G$ is the gravitational constant. 
If the QmCl undergoes the gravitational collapse it is no more useable 
as a device to measure time intervals because the products of the decays
({\it e.g.} the photons of the example discussed above) 
cannot escape outside the Schwarzschild radius to bring the information 
that time is flowing in that region of space. 
This implies that the smallest radius of the QmCl is its Schwarzschild
radius, $R \ge R_{\rm Sch}$ or 
\begin{equation}
\label{eq:radius}
\frac{1}{M} \ge \frac{2 G}{c^2 R}
\end{equation}
The condition above has been discussed in the literature as a necessary
lower limit on the size of a massive clock. In particular Amelino-Camelia
proposed an equation for a lower bound on the uncertainty in the measurement
of a distance in which the condition above is included (Amelino-Camelia, 
G., 1994, Mod. Phys. Lett. A9, 3415, 1996, Mod. Phys. Lett. A11, 1411). 

Inserting (\ref{eq:radius}) into (\ref{eq:3clock}) gives
\begin{equation}
\label{eq:1uncertainty}
\Delta t R \ge \frac{1}{\epsilon^2}
\frac{G \hbar}{c^{4}},
\end{equation} 
where $R = \Delta r$ is the radius of the QmCl 
({\it i.e.} the uncertainty on the exact position of the radioactive
decay).
Because, as we noted in \S \, \ref{quantumclock1}, $\epsilon \le 1$, 
we can write 
\begin{equation}
\label{eq:spacetime}
 \Delta r \Delta t \ge \frac{G \hbar}{c^{4}} 
\end{equation}
The equation above quantifies the impossibility to simultaneously 
determine spatial and temporal coordinates of an event with
arbitrary accuracy.

In order to demonstrate that relation (\ref{eq:spacetime}) holds 
independently of the kind of clock adopted, we discuss, in the Appendices,
different kinds of QmCls, based on different, although fundamental, quantum 
processes, namely the Blackbody emission (Appendix A) and the 
Hawking radiation (Appendix B).

\section{Discussion}
\label{discussion}

In the framework of String Theory a simple space-time uncertainty relation
has been proposed which has the same structure of the uncertainty relation 
discussed in this paper (Yoneya T. 1987, p. 419 in ``Wandering in the Fields'', 
eds. K. Kawarabayashiand A. Ukawa, World Scientific; Yoneya T. 1989, Mod. 
Phys. Lett. A4 1587; see {\it e.g.} Yoneya T. 1997, arXiv:hep-th/9707002v1 
for a review). The relation proposed in String Theory constraints the 
product of the uncertainties in the time interval $c \Delta T$
and the spatial length $\Delta X_l$ to be larger than the square of the 
string length $\ell_S$, which is a parameter of the String Theory. 
However, to use the same words of the proposer, this relation is 
``speculative and hence rather vague yet''. 

On the other hand, in this paper we demonstrate that an uncertainty relation 
between space and time is a necessary consequence of the well known result 
of GR that gravitational collapse is unavoidable once a given amount of 
mass-energy is concentrated into a spatial extension smaller than the volume 
encompassed by the Event Horizon of that amount of mass-energy.
In other words we show how an uncertainty relation between space and time is
necessarily implied in the framework of {\it any} fundamental theory
once the Uncertainty Principles of Quantum Mechanics and the existence 
of Event Horizons (as predicted by GR) are properly taken into account. 
Indeed, in the context of Field Theories, uncertainty relations 
between space and time coordinates similar to that proposed here
have been discussed as an ansatz for the limitation arising in combining 
Heisenberg's uncertainty principle with Einstein's theory of gravity
(Doplicher S., Fredenhagen K., Roberts J. E., 1995, Commun. Math. Phys.,
172, 187).
The relation proposed here does not depend 
on parameters defined within a specific theory but only on fundamental 
constants, since the String Theory parameter $\ell_S^2$ is replaced by
$G \hbar/c^3$. 

Here we briefly outline some of the implications of the uncertainty
relation between space and time derived above. 
To this aim it is useful to represent space and time intervals 
in a standard space-time diagram.
We choose the space and time units in order to have $c =1$, or 
$c \Delta t$ as the ordinate. In this representation the
bisector defines the null intervals, separating the timelike intervals,
above the bisector, from the spacelike intervals, below.
The relation (\ref{eq:spacetime}) of \S \ref{relativity}, namely
\begin{equation}
\label{eq:2spacetime}
 \Delta r \, c \Delta t \ge \frac{G \hbar}{c^{3}} 
\end{equation}
defines an hyperbola in this plane whose asymptotes are the $\Delta r$
and $c \Delta t$ axes and whose vertex is located at 
$(\Delta r)_{\rm vertex} = (c \Delta t)_{\rm vertex} = \sqrt{G \hbar/c^3}$. 

The following considerations can be made: \\
i) The minimum (measurable) space distance is the Planck Length. 
This is because a proper space distance is defined 
for spacelike intervals and the minimum ``$x$'' coordinate of the points 
below the bisector is $(\Delta r)_{\rm min} = \sqrt{G \hbar/c^3}$, which
is the Planck Length. \\
ii) The minimum (measurable) time interval is the Planck Time. 
This is because a proper time interval is defined 
for timelike intervals and the minimum ``$y$'' coordinate of the points 
above the bisector is $(\Delta t)_{\rm min} = \sqrt{G \hbar/c^5}$, which is
the Planck Time. \\
iii) The uncertainty relation is invariant under Lorentz transformations.
This is because $(\Delta r)' = \gamma^{-1} \Delta r$ 
(Lorentz contraction) and $(\Delta t)' = \gamma \ \Delta t$
(time dilation), where $(\Delta r)'$ and $(\Delta t)'$ are space and time 
intervals measured in a reference frame moving at uniform speed $v$
with respect to the laboratory frame, and $\gamma = (1 - (v/c)^2)^{-1/2}$
is the Lorentz factor.

Relation (\ref{eq:2spacetime}) has the same form of the uncertainty 
relations (\ref{eq:heis}) and (\ref{eq:heis2}) which hold, in QM,
between non-commuting quantities. 
This relation has been obtained as an
unavoidable logical consequence of the very first principles of the
QM (the Heisenberg uncertainty relations) and of the GR (the formation
of an Event Horizon during gravitational collapse). In this respect 
{\it any} theory that will consistently describe the behavior of matter
on the smallest spatial and temporal scales under conditions in which
gravitational effects are not negligible {\it has to} take into account
relation (\ref{eq:2spacetime}). It is thus reasonable to consider this
new uncertainty relation as one of the fundamental principles over which
Quantum Gravity has to be built. Indeed, it is possible to develop QM 
adopting the Heisenberg uncertainty principles (and the commutation
relations associated) as the postulates over which the whole quantum 
theory is built. In a similar way, it looks possible to develop 
the foundations of a mathematical theory of gravity which will be fully 
consistent with the postulates of the quantum theory, starting from the
uncertainty relation between space and time discussed above. 

Although relation (\ref{eq:2spacetime}) has the structure of an uncertainty
relation, and therefore does not contain a minimum spatial length or a 
minimum time duration explicitly, the timelike and spacelike classification
of the intervals, determined by Special Relativity, when combined with 
(\ref{eq:2spacetime}), implies (in a somewhat unexpected way) the 
existence of minimal space-time ``quanta'' equal to the product of the Planck 
Length and Time, respectively. 
In other words, this new uncertainty principle naturally implies the 
existence of ``atoms'' of space and time whose size does not require the 
introduction of any extra parameter in the theory. This quite remarkable
feature arises as the unavoidable logical extrapolation resulting from
the combination of the Uncertainty Principles of QM, with the 
universal hyperbolic character of the metric implied by SR, and with the 
Space Closure occurring, according to GR, during complete gravitational 
collapse: QM, SR, and GR enter in the uncertainty relation through their 
fundamental constants, $\hbar$, $c$, and $G$. We finally note that in the 
limits $\hbar \rightarrow 0$ and $c \rightarrow \infty$ (that means in the
classical limit), there is no uncertainty relation between space and
time, as expected. A similar discussion on the role of $\hbar$, $c$, and $G$ 
in determining a spatial resolution limit can be found in 
Garay (1995, Int. J. Mod. Phys. A10, 145).

\section{Supplementary Materials}
\subsection{Another kind of Quantum Clock: the Blackbody Clock}
\label{quantumclock2}

A ``Blackbody Clock'' (BBCl, 
hereafter) is made of a box 
where a given amount of matter is
in thermodynamic equilibrium with its electromagnetic radiation. 
In this case the radiation spectrum is
a blackbody.
To describe this BBCl in a more quantitative way,
let us suppose a spherical box of radius $R$ and a blackbody of 
temperature $T$. In this case the luminosity of the blackbody is
$L = 4 \pi R^2 \sigma T^4$, where $\sigma = a c /4$ is the Stefan 
Boltzmann constant, $a = (8 \pi^5 k^4)/(15 c^3 h^3)$ is the radiation
(density) constant, and $k$ is the Boltzmann constant. 
Given that the mean photon energy is 
$<h\nu> \; \sim 3kT$, the photon emission rate is 
\begin{equation}
\label{eq:1bbclock}
\frac{dN_{\rm ph}}{dt} = \frac{4\pi R^2\sigma T^4}{3kT} =
\frac{4}{3}\pi R^3 a T^4 \times \frac{c}{4RkT} =
E_{\rm BB} \times \frac{c}{4RkT} 
\end{equation}
where $E_{\rm BB} = M_{\rm BB}c^2$ is the energy of the blackbody.
A short time interval of duration $\Delta t$ can be measured by the BBCl
by simply counting the number of photons detected 
${\cal N}_{{\rm ph}\Delta t}$. The detection is a quantum process subject
to the counting (Poisson) statistics, thus the relative error on this
measure is $\epsilon = \delta t / 
\Delta t \sim {\cal N}_{{\rm ph}\Delta t}^{-1/2}$ or
${\cal N}_{{\rm ph}\Delta t} \sim \epsilon^{-2}$.
Therefore we have
\begin{equation}
\label{eq:2abbclock}
\Delta t \times dN_{\rm ph}/dt = {\cal N}_{{\rm ph}\Delta t} 
= \epsilon^{-2} 
\end{equation}
and, expressing $dN_{\rm ph}/dt$ with (\ref{eq:1bbclock}) and
the blackbody energy through its mass we find
\begin{equation}
\label{eq:2bbclock}
\Delta t = \frac{1}{\epsilon^2 M_{\rm BB}c^2} \times \frac{4RkT}{c}
\end{equation}
Note that (\ref{eq:2bbclock}) and (\ref{eq:2clock}) of 
\S \, \ref{quantumclock1} are identical providing that 
$M \rightarrow M_{\rm BB}$
and $E_{\mathrm p} / \lambda \rightarrow 4RkT/c$. 

In \S \, \ref{heisenberg} we used the time-energy uncertainty relation to
derive $E_{\mathrm p} / \lambda = E_{\mathrm p} \tau \ge
\delta E_{\mathrm p} \times \delta t \ge \hbar / 2$. Here we note that $3kT/c = 
<h\nu>/c = <p_{\rm ph}>$, where $<p_{\rm ph}>$ is the modulus of the mean
momentum of the emitted photons. Since all the photons are emitted from
a region of size $\le R$, we know their position along the 
radius of emission with an uncertainty $\delta r \sim R$. Thus we have
\begin{equation}
\label{eq:3bbclock}
\frac{4RkT}{c} = \frac{2}{3}\;\; R \times <p_{\rm ph}>  
\ge \frac{2}{3}\;\; \delta r \times \delta <p_{\rm ph}>_r 
\ge \frac{2}{3} \;\; \frac{\hbar} {2}
\end{equation}
where the last inequality follows from the position-momentum uncertainty
principle in a given direction $r$:
\begin{equation}
\label{eq:heis2} 
\delta r \times \delta p_r \ge \hbar / 2
\end{equation}
and where we used the
obvious fact $<p_{\rm ph}> \ge \delta <p_{\rm ph}>_r$\footnote{
The fact that there is a lower limit in the momentum of photon whose 
position is constrained to be within a given region of space (in the 
same direction of the momentum) is a direct consequence of the Heisenberg 
uncertainty relation between momentum and position of a particle.
This behavior is explained in the framework of Classical Electrodynamics 
by the fact that there is a maximum wavelength allowed for standing waves 
in a cavity, which is twice the cavity size, once the correct boundary 
conditions (the electric field should vanish at the walls of the cavity) 
are taken into account. 
A similar relationship holds in QM for the so called ``zero point'' energy 
(and thus momentum) of a particle confined in a box. 
An interesting experimental proof of this behavior for photons confined 
in space has been given by Hulet, Hilfer, and Kleppner (Hulet R. G., 
Hilfer E. S., and Kleppner D. 1985, Phys. Rev. Lett., 55, 2137).
Spontaneous radiation emission by an atom in a Rydberg state is observed
to turn off abruptly once the size of the cavity in which the atom has 
been confined is reduced below half of the wavelength of the photon that
should be emitted in the spontaneous decay of the Rydberg atom. Indeed 
application of the Heisenberg uncertainty relation between momentum and
position of a particle, to a photon with a momentum along a given direction
$X$, gives $\Delta X \ge \lambda_{\rm ph}$, where $\lambda_{\rm ph}$ is the
wavelength of the photon. If the photon wavelength is constrained by the
presence of the cavity to be smaller than $\lambda_{\rm ph}$, in order not  
to violate the Heisenberg uncertainty relation, its momentum must be larger
than $\hbar /\lambda_{\rm ph}= \hbar \nu_{\rm Ry}/c$, where $\nu_{\rm Ry}$ 
is the frequency associated with the transition of the Rydberg atom. Since
the transition energy (and therefore its frequency) is fixed by the atomic
structure, spontaneous emission is inhibited by the quantum uncertainty
principle.}. 

Inserting (\ref{eq:3bbclock}) in
(\ref{eq:2bbclock}) we find:  
\begin{equation}
\label{eq:4bbclock}
\Delta t  \ge \frac{\hbar}{3 \epsilon^{2} c^{2}} \times \frac{1}{M_{\rm BB}}
\end{equation}
which is similar to eq. (\ref{eq:3clock}) of \S \, \ref{heisenberg}.
Since the BBClk is made of matter and radiation in thermodynamic 
equilibrium, we have $M = M_{\rm matter} + M_{\rm BB} \ge M_{\rm BB}$ 
or $1/M_{\rm BB} \ge 1/M$, where $M$ is the whole mass of the 
clock\footnote{Given enough time for initial thermalization, 
$M_{\rm matter}$ can be reduced to an arbitrarily small amount of
mass, which means $M \rightarrow M_{\rm BB}$ in this case.}.

We finally include the limits imposed by the General Relativity 
discussed in \S \, \ref{relativity} by inserting (\ref{eq:radius}) 
in (\ref{eq:4bbclock}):
\begin{equation}
\label{eq:5bbclock}
\Delta r \Delta t \ge \frac{2}{3 \epsilon^2}
\frac{G \hbar}{c^{4}}
\end{equation}
Because, as we noted in \S \, \ref{quantumclock1}, 
$\epsilon \le 1$, assuming (as a lower limit) $\epsilon \sim 1$
and neglecting the factor $2/3$, 
we find that the uncertainty relation between space and time
(equation (\ref{eq:spacetime}) of \S \, \ref{relativity}) holds for
BBCl.  

\subsection{The extreme Quantum Clock: the Hawking Clock}
\label{quantumclock3}

In this section we briefly discuss a BBCl whose source of particles
is the Hawking--Beckenstein radiation emitted by the Event Horizon 
of a Black Hole as theoretically demonstrated by Hawking 
(Hawking S. W., 1975, Commun. math. Phys., 43, 199). In the
following we refer to this kind of clock as the Hawking Clock (HwCl).
Again we consider the rate of particles emitted by this clock as in 
(\ref{eq:1bbclock}): 
\begin{equation}
\label{eq:1hwclock}
\frac{dN_{\rm part}}{dt} = \frac{4\pi \sigma }{3k} R_{\rm BH}^2 T_{\rm BH}^3 
\end{equation}
where $R_{\rm BH} \sim R_{\rm Sch} = 2GM/c^2$ and 
$T_{\rm BH}= \hbar c^3/(8 \pi kGM)$ are the Event Horizon radius and the Black
Hole temperature, respectively, and $M$, here, is the Black Hole mass.
Using equation (\ref{eq:2abbclock}) as in \ref{quantumclock2} we get 
\begin{equation}
\label{eq:2hwclock}
\Delta t  = \epsilon^{-2} \times \left( \frac{dN_{\rm part}}{dt} \right)^{-1}
\end{equation}
The size of HwCl, $\Delta r$, cannot be less than the radius of 
the Event Horizon of the Black Hole 
\begin{equation}
\label{eq:3hwclock}
\Delta r \ge R_{\rm BH}
\end{equation}
Multiplying equations (\ref{eq:2hwclock}) and (\ref{eq:3hwclock}) and
inserting equation (\ref{eq:1hwclock}) we get
\begin{equation}
\label{eq:4hwclock}
\Delta r \Delta t \ge \epsilon^{-2} \frac{3k}{4\pi\sigma} 
(R_{\rm BH}T_{\rm BH}^3)^{-1} = 2^8\,3^2\,5\,\epsilon^{-2} \frac{(GM)^2}{c^5} 
\end{equation}
which shows that, for the HwCl, the uncertainty relation between space and 
time depends on the square of the Black Hole mass, its minimum occurring for
the smallest possible mass for a Black Hole undergoing emission of Hawking
radiation. Since such Hawking Black Hole radiates as a Blackbody with mean
particle energy $<E_{\rm part}> \; \sim 3 k T_{\rm BH}$, 
we assume that its minimum
mass occurs when the (mean) particle emitted carries away all the Black Hole
energy $M_{\rm min}c^2$, namely $<E_{\rm part}> \; \sim 3 k T_{\rm BH} = 
M_{\rm min}c^2$ which gives
\begin{equation}
\label{eq:5hwclock}
M_{\rm min} = \left(\frac{3}{8 \pi}\right)^{1/2}
\left(\frac{\hbar c}{G}\right)^{1/2} 
\end{equation}
which, beside the factor $\sqrt{3/(8 \pi)}$, is the Planck Mass. Inserting
(\ref{eq:5hwclock}) in (\ref{eq:4hwclock}) we get
\begin{equation}
\label{eq:6hwclock}
\Delta r \Delta t \ge \epsilon^{-2} \times \frac{2^5\,3^3\,5}{\pi} 
\times \frac{G \hbar}{c^4} 
\end{equation}
Because, as we noted in \S \, \ref{quantumclock1}, 
$\epsilon \le 1$, assuming (as a lower limit) $\epsilon \sim 1$
and neglecting the factor $2^5\,3^3\,5/\pi > 1$, 
we find that the uncertainty relation between space and time
(equation (\ref{eq:spacetime}) of \S \, \ref{relativity}) holds for
the HwCl.

\bibliographystyle{elsarticle-harv}
\bibliography{<your-bib-database>}






\end{document}